# Exploring the Role of Awareness, Government Policies, and Infrastructure in Adapting B2C E-Commerce to East African Countries Tanzania Case Study

Emmanuel H. Yindi, ImmaculateMaumoh and Prisillah L. Mahavile

**Abstract:-**It has considered almost 30 years since the emergence of e-commerce, but it is still a global phenomenon to this day. E-commerce is replacing the traditional way of doing business. Yet, expectations of sustainable development have been unmet. There are still significant differences between online and offline shopping. Although many academic studies have conducted on the adoption of various forms of e-commerce, there are little research topics on East African countries, The adoption of B2C e-commerce in East African countries has faced many challenges that have been unaddressed because of the complex nature of e-commerce in these nations. This study examines the adaptation of B2C in East Africa using the theory of diffusion of innovation. Data collected from 279 participants in Tanzania were used to test the research model. The results show that awareness, infrastructure innovation and social media play a significant role in the adoption of e-commerce. Lack of good e-commerce policy and awareness discourages the adoption of B2C. We also examine how time influences the adaptation of B2C e-commerce to the majority. So, unlike previous adoption studies, which have tended to focus on technology, organizational, and environmental factors, this study guides the government on how to use social media to promote B2C e-commerce.

## I. INTRODUCTION

Today, the Internet has become an indispensable part of people's lives. Approximately people all over the world recognize that the Internet plays an important role in our lives and has led to the production of employment opportunities and developments in business, especially B2C e-commerce.(Wang, Lu and Commerce 2014). Growth of e-commerce has been driven by the rapid adoption of technology, Increasing use of devices such as smartphones, tablets, broadband Internet access, 3G, 4G and the credibility of e-commerce companies that have led to an increase in online customer base(Rahayu and Day 2017)It should be marked that B2C e-commerce is not a whole new concept, but it has seen increasing and unpredictable developments in recent years. Indeed, the Internet can be seen as a major justification for the development of e-commerce. But this advantage has not yet been realized in East Africa Countries, although the numbers of Internet users in Tanzania increased by 16% at the end of 2017 to reach a total of 23 million, the majority of them use their devices to connect for social purposes and not to online businesses (ZoomTanznaia2017).According to(Paris, Bahari and Iahad 2016), B2C e-commerce (business to consumer electronic commerce) is used to describe a business transaction between a company and a consumer. Traditionally, the term referred to the process of selling products directly to consumers, including in-store purchases or eating in a restaurant. Currently, it describes transactions between online retailers and their customers.Tanzania, officially the United Republic of Tanzania, is an East African country in the Great Lakes region of Africa, has the largest population in East Africa estimated at 59,734,218 (UN data 2019) total area is 947,300 km2 that covers an area about twice the size of California. (Nkwabi et al. 2018). The largest city of Dar es Salaam with a population of 6.7 million is located along the east coast on the Indian Ocean. Tanzania's current GDP was USD 63 billion in 2019, making it the second-largest economy in East Africa after Kenya and the 6th largest in sub-Saharan Africa(Group 2018). Based on the Tanzanian population, the total number of Internet users is about 14.72 million, while internet penetration is about38.7%.(Internet world stats 2019), mobile phone penetration is 83%; 3G/LTE services (2018), while the total number of mobile cellular subscribers is about 43,953,860 (CIA World Factbook 2020); according to a general assessment, telecommunications services are marginal. The huge population and vast and fragmented geography provide a good reason for businesses in Tanzania, especially B2C e-commerce. But for now, Tanzanians are still lagging in the adoption of e-commerce. In e-commerce index, Tanzania ranks second in east Africa, 16th in Africa and 110th in the world (UNCTAD e-commerce2018). Cash on delivery remains the most preferred method ofpayment, and 94% of Tanzanians opt for cash on delivery as their preferred method of payment, which is to declare B2C adaptation in Tanzania still a challenge. While Internet plans are becoming more affordable, some people in Tanzania still do not want to use their data for anything other than WhatsApp, Facebook, and checking their emails. However, the reality is that searching for products online does not use a lot of data (if you have a package) and is cheaper than taking a bus or using fuel to drive to a store location (Zoom Tanzania 2017).(Park and Kang 2014) Found that the government should play a role that facilitates the development of e-commerce by providing strong secure online payment options, ensuring a strong ICT infrastructure, providing educational programs and raising awareness through a variety of means such as media and academic institutions(Awiagah, Kang and Lim 2016). B2C growth depends on expanded and affordable access to critical infrastructure through technology integration, telecommunications policy, strong network infrastructure, adequate bandwidth and support for target applications. (Thomas, Jose and Technology 2015) one of the dominant factors for which the adaptation of e-commerce in developing countries is lagging is the lack of infrastructure, illiteracy, slow Internet connections, non-progressive monopoly of the sector and the issue of awareness. Rising awareness remains a key factor in the adoption of e-commerce. Awareness will contribute positively to improving the prevalence of e-commerce and increasing the number of Internet users.(Al-Khaffaf 2013) The B2C Awareness Study is important because it can reduce customer uncertainty and concerns, increasing consumer confidence. Consumer awareness remains a significant impact on the adoption of e-commerce. (Nathan et al. 2019)there is a constructive relationship between awareness and the adoption of e-commerce. Showing the benefits of knowledge about e-commerce can increase the level of adoption. However, newspaper analysis shows that there is a lack of awareness in Tanzania. (Kabango, Asa and Development 2015)and (Kabir, Musibau and Management 2018) both stated that lack of awareness in developing countries is one of the major obstacles to the adoption of B2C. The dissemination of information through social media and integration could play an important role in raising awareness of e-commerce.(Hajli and Management 2012). Today, more than 90% of the online adult population uses social media, which includes many of your customers that are in contact with other people. Global statistics show that while the average daily time spent online in 2012 was around 1 hour and 15 minutes, it is now closing to 2 hours, 45% of which is spent on social media and micro-blogging sites. (Taprial and Kanwar 2012). No company or country can ignore social media any longer. If you do, not only do you lose an opportunity to improve your business, but your absence can also damage your brand and reputation. Social influence has always played a major role in customers' buying decisions. But today, more and more people are relying on online social media to get references, and recommendations from others, ask/answer questions and share their experiences. Companies generally refer to social media as consumer-generated media (CGM). Social media can be differentiated from industrial or traditional media, such as magazines, newspapers, television and cinema, as they are relatively inexpensive, easily accessible and allow anyone (individuals) to publish or access information. Social media represents the future of e-commerce, these media have a huge impact on Internet users, especially on the way people communicate and exchange information via commonly used sites such as Facebook, My Space and YouTube(Abed, Dwivedi and Williams 2015)Social media is seen as a solution for e-commerce (Hajli and Management 2012)People can be reached very easily via social media, and e-commerce opportunities can be introduced. Therefore, the government should take a step forward in using social media to increase the level of public awareness; this will simplify the acceptance of the concept of e-commerce in the future (Ahmad et al. 2015) An academic researcher unanimously confirmed thatawareness was one of the crucial issues directly affecting the development of e-commerce. Besides, the media has a significant impact on improving knowledge and serves as a conduit for the government to use and increase awareness. The world's fastest-growing economies by 2030 will be in Africa. This makes thecontinent the next major e-commerce market(Leke, Chironga and Desvaux 2018)and as this positive narrative continues to place Africa as the highest investment destination, the need for modern infrastructure systems has become inevitable. The growth of e-commerce will depend to a large extent on the quality and efficiency of logistics networks, such as national address systems and appropriate road networks. (ISMAIL 2020). Popular media platforms like Facebook, Twitter and Instagram which have around 20 million users in Tanzania might be applied as a tool of advertisement, and raise knowledge. The millions of customers' are there just waiting for you or us to launch your products or services on social media pages, also social media can be used to lift awareness and trust. Last, this social study will be encouraging the adaptation of B2C e-commerce even though is not a simple game, hard work is needed to attain achievement, therefore, it's not an overnight success story although that can happen.

## II. THEORETICAL BACKGROUND

Many technologies are launched to the market every day. To understand how these technologies are being adopted on the market, we need to use the theory of diffusion of innovation formulated by (Rogers and Diffusion 1962). We need to understand that this theory is as relevant today as it was then.(Rogers 1976)Analyze how social members (the members or units of a social system can be individuals, informal groups, organizations or sub-systems) adopt new innovative ideas and how they have made the decision to adopt them. According to Roger, the media and interpersonal communication channels are both involved in the diffusion process. (The communication channel is how messages are passed from one individual to another, for example, through a word of mouth, SMS or other forms of literature, etc.). Innovations should be adopted to achieve development and sustainability(Sahin 2006). Also(Rogers 1976)proposed that time remains a crucial aspect in the adaptation process, the length of time needed to decide on the adaptation of innovation to use new ideas is also relevant when studying how an individual or other adoption unit changes its internal state (e.g. knowledge or decision to adopt) and its manifest behaviour (actual adoption or rejection). Furthermore, according to (Rogers 1976), time is a significant measure when we classify adopters into different categories. (Agrawal and Zeng 2015) consider that mobile phones have taken a long time to spread among people after they have been introduced to the market. Who decided to accept innovation? (Rogers 1976) proposed that in a social system, decisions are made in three ways. He suggested three ways of taking into account people's ability to make judgments of their own accord, and their ability to implement them voluntarily; these three ways are as follows. First opinion — individuals decide innovation in the social system on their own, a second collective — the decision made by all individuals in the social system, and a third authority — few individuals decide for the social system as a whole (Hwang and Lin 2012). Furthermore, Rogers identifies the mechanism of diffusion of innovation theory through five, stages(Sahin 2006) that are i. knowledge, an individual may naturally expose the innovation, but shows no interest in it due to lack of information or awareness of the innovation (Ritala et al. 2015) ii. Persuasion, once a consumer hears about an innovative product, the company's current goal is to persuade him to use and buy it. After accumulating knowledge (e.g. a discussion with a colleague) and being intrigued by a product, the consumer is ready to be persuaded. iii. The decision, at this stage, an individual analysis, the pros, and cons of the innovation and decide to accept/reject the innovation. (Rogers 2010) explains, "One of the most difficult steps in identifying the evidence" iv. Implementation even when the consumer decides to make the purchase, the company still has to convince the consumer that the product is useful, and may still need to be educated about the most excellent features of the product to maximize the product engagement experience(Kim 2020).The rate of adoption is also influenced by the social system in which an innovation spreads. (Rogers 2010) Mention weak ties, opinion leaders, social learning and critical mass as important concepts that help understand the diffusion of innovations through social networks.

*2.1 Awareness*
 According to (e Silva et al. 2020) the awareness campaign is not particularly important because it draws attention to the subject of B2C. It is also a part of the broader strategy to motivate people to take action; it is an essential first step in bringing about societal change and creating positive changes in culture. The awareness of e-commerce influences trust, consciousness continues to be a critical element of the adoption of B2C in Tanzania (Kabanda, Brown and Informatics 2017) without the right knowledge of e-commerce adoption cannot be achieved. Consciousness could be enhanced by social media. Awareness can also increase by adopting adequate training and education in schools and universities. Thus, it is the role of the government to provide training and education for e-commerce (Yaseen, Dingley and Adams 2015)Factors hindering the adoption of e-commerce in Tanzania include the lack of regulation and legislation; lack of awareness and education on B2C e-commerce and trust issues (Park and Kang 2014) Other studies of B2C e-commerce is (Najafi 2014) who argued that awareness remains the first step towards trust; consequently, the first step towards fostering e-commerce in developing countries, and there is a positive relationship between awareness and adoption of e-commerce. (Kabango et al. 2015) The use of B2C services, which have provided a good example of e-commerce, obtains new knowledge to many customers in Tanzania, and the lack of consciousness on online business remains a crucial factor preventing customers from adopting it. (Kurnia et al. 2015)Declared the lack of consciousness, and knowledge of B2C contributes to the challenges of e-commerce adoption. There is a constructive relationship between e-commerce awareness and adoption, so e-commerce awareness is not avoided because it creates a spark for B2C adaptation. ((Gebert-Persson et al. 2019)) As knowledge increases, customer confidence in technology and enterprise increases. So, we argue that the more knowledgeable of the customers the more trusting attitude increase. According to (Mayer, Davis et al. 1995) Characteristics and actions of the trustee will lead that person to be more or less trusted. Awareness campaigns should be disseminated on the most appropriating channels to contact audiences targeted not only through traditional channels but also through other forms of communication, such as social media (Oláh et al. 2019).Arguing furthers at some point campaigns need to go beyond raising awareness instead of trying to get things done. Any organization must realize it comes a time when they need to stop raising awareness about e-commerce instead inspire people to take action about that. The government needs to have a theme behind what it sells before they can create a strategy because the theme helps a government better recognize what it is trying to promote and consequently who will be most interested in what the trying to sell.(Turban et al. 2017) offered three main reasons why some small and medium-sized developing enterprises still reluctant to adopt e-commerce technology, these include, the lack of success stories, the lack of knowledge of e-commerce, and the lack of information on the potential impact of e-commerce implementation on business performance. Developing awareness is not only to increase awareness of the B2C service but also it increases the trust in e-commerce(Chandra and Kumar 2018). Organizations often believe that by simply raising awareness, people will interact with the cause and change their attitude, but in a real sense, this is wrong because without undertaking the necessary campaign against the awareness message may not be delivered and is possible for not achieving an expected goal or target.(Abed et al. 2015)

*2.2 Social Media*
 The emergence of social media dates back to the early days of the Internet when people began to share information and communicate with each other. Social media are all media or platforms that allow people to be social or become social online by creating and sharing content, news, photos, videos, etc. with other people. (Taprial and Kanwar 2012) If we separate the two terms: the term social refers to the interaction with other people, and the sharing or receiving of information, while the term media refers to the means of mass communication, which in the traditional sense include television, radio and newspapers(Taprial and Kanwar 2012). So, all web applications that enable the creation/exchange of user-generated content and interaction between users can be classified as social media. These may include social networking sites (Facebook, Twitter, Google+), blogs, internet forums, bookmarking sites, online community sites, mobile question and answer and messaging sites chat applications etc. (Hiebert and Gibbons 2017). The number of social media user in Tanzania is increasing significantly everyday Internet penetration in this country of around 60 million people had increased to 54 % in 2017 from 49% a year before but most users of social media are for leisure, citizens have not yet transferred their online time for e-commerce activities. The reasons for this may be that people are still not aware of the real benefits of adapting to B2C e-commerce. (TCRA Report 2017) Meanwhile (Abed et al. 2015).Argued that Social Media as a Bridge to the adaptation of e-commerce. Online forums, web blogs, podcasts and popular media are rapidly becoming an essential part of the daily, personal, social, and business life as well as a major source of customer information and a channel of communication, information sharing, and distribution of creativity and Individual entertainment(Cawsey, Rowley and Planning 2016) It is a simple and inexpensive solution to reaching potential customers, listen to the voices of customers, creating vast business networks, (Glenn 2013)  Social media offer organizations and customers new ways to connect them. Businesses have begun to embrace social media websites as a way to improve information sharing, communication and collaboration by implementing many innovative and essential business practices(Iankova et al. 2019). Kenyans are traditionally wary of online payment 94% of online shoppers choose cash on delivery as the method of payment and prefer to have the free return option (Gachenge 2020). Lacks appropriate delivery addresses and trust in sellers has also hampered the growth of e-commerce. (Park and Kang 2014) using mass media is possible for the government to change the behaviour of the population of e-commerce and building trust. Today, a significant percentage of advertising campaigns take place through social media sites. Many social media users shop online after sharing articles on Pinterest, Facebook and Twitter. This is a clear indication of how essential social networks can stimulate adaptation to B2C e-commerce (Leong, Jaafar and Ainin 2018). Meanwhile  (Park and Kang 2014)believed that National Policy Initiatives (NPI) and the government action, for example, training, advertising on mass media are helping to promote B2C e-commerce in developing countries. Social media has a profound effect on the world in a short time by connecting people and businesses in new and exciting ways. Now, social media has an impact on general e-commerce, besides that is used by brands as a way to advertise, increase their online presence and providing high-quality customer service. According to (Beier and Wagner 2015) pointed out that media, such as newspapers, television and radio can be used as a tool to broadcast extra knowledge and awareness about e-commerce in developing countries. Through social networks, people share common interests or needs that did not normally meet; they support each other in knowledge, sharing and solving problems (Zhang et al. 2017). Promotion of B2C in popular social media is another method that can be used to raise awareness of online business and B2C in general(Riu and Review 2015).

*2.3 Infrastructure*
　　The performance of B2C in developing countries, particular Tanzania, faces significantly greater challenges than B2C in developed countries due to unreliable Internet connection, which is unsatisfactory because of unstable infrastructure, high maintenance costs and the basic level of ICT penetration throughout the country (Kabango et al. 2015)While (Park and Kang 2014) Suggested adaptation of B2C among Ugandans facing challenges due to environmental factors such as unreliable as grid infrastructure and electricity. The government faced challenges to develop good strategies and policies that will strengthen infrastructure to encourage e-commerce. (Oreku et al. 2013) Argued the role of government in developing countries remains important in facilitating the essential conditions for the development of e-commerce, like providing strong secure online payment options, ensuring a strong ICT infrastructure, delivering educational programs and raising awareness through such means as media and educational institutions (Chaffey, Edmundson-Bird and Hemphill 2019)meanwhile (Gorla, Chiravuri and Chinta 2017) showed how the level of e-commerce knowledge within an organization can influence its adoption? They pointed out that knowledge among non-IT professionals is the most critical factor in Internet adoption. While (Alrousan and Jones 2016) argued that e-commerce adopting in SMEs in developing states is slow compared to developed countries due to many obstacles, like ICT infrastructure, level of economic, cultural, legal, and social, computer skills among people and postal infrastructure factors. (Oreku et al. 2013)Private organizations are not organized enough to provide an IT infrastructure that the executive should launch programs to reduce these barriers. Most African countries road structure is organized non-properly it is characterized by heavy traffic jams that lead to delayed deliveries, cancelled orders for on-demand services and therefore, loss of revenue.(Alyoubi 2015) The temporary method to meet this infrastructure challenge in the delivery process is using motorcycles, popularly known as Bodabodas. About 1.2 million passenger motorcycles in Kenya and Tanzania are used by online companies in the B2C service these companies have sought to tap into online markets using Bodabodas to deliver products quickly, especially in active cities. (Biztech Africa 2019) Governments of developing country should create a favorable policy and appropriate environment for e-commerce and ensure a strong infrastructure, such as the availability of reliable and cost-effective broadband and verified and maintained roads all the time. (Bilgihan et al. 2016) While (Di Castri and Gidvani 2014) argued that although e-commerce in Tanzania is still at the embryonic level of development, it can grow rapidly, especially after considering the recent expansion of mobile payment methods such as M-Pesa, TigoPesa, Airtel Money and Zantel Money.

## III. RESEARCH MODEL AND HYPOTHESES

*3.1 Awareness*

Awareness remains a key factor in the adoption of e-commerce. Awareness-raising is a positive contribution to increase the prevalence of e-commerce and increase the number of The Internet users (Al-Khaffaf 2013) The study of B2C awareness is important because can reduce uncertainty and customer concerns, and thus, increase trust in customers. The main barriers to e-commerce adoption in Tanzania are due to lack of knowledge, lack of e-commerce skills and lack of skilled workers to operate e-commerce (Kabanda et al. 2017). According to (Beier and Wagner 2015) pointed out that media, such as newspapers, television and radio can be used as a tool to broadcast extra knowledge and awareness about e-commerce in developing countries.(Thomas et al. 2015)Social networking services allow clients to share their experience with friends, receive their recommendations, opinions, advice and communication. In recent years, we can see that social media plays a very important role in driving sales and increasing the popularity of e-commerce. While (Chong et al. 2014) stressed that effective knowledge processes like knowledge acquisition, dissemination, and application are important for the adoption of new technologies. Knowledge dissemination is a transfer and sharing of knowledge by the enterprise (Masa'deh et al. 2015). The primary aim of knowledge dissemination is to distribute knowledge to all those involved in a particular business process. And knowledge application includes the assimilation of the knowledge generated at the acquisition and dissemination levels that applied in the routine.

- *H1a Awareness has a positive effecton B2C adaptation.*
- *H1b Social Media has a positive effect on the awareness to adapt B2C e-commerce*
- *H1c There is a significant and meaningful relationship betweenthe length of time and interpersonal communication in the adaptation of B2C e-commerce.*

*3.2 Social Media*

Social media and mainstream media play a significant role in raising awareness of e-commerce. So, these media remain the future of e-commerce(Han and Kim 2016) Social media have the largest impact on The internet users, especially on how people communicate and exchange information through common sites such as Facebook, My Space, and YouTube (Abed et al. 2015). Social media are seen as a solution for trust and adoption of e-commerce. (Makki and Chang 2015), while roger stated that both mass media and interpersonal communication channels involved in the dissemination process. Popular media are no longer sites where people can access and exchange information. Social media serve as a market strategy by facilitating the process of getting customers, attracting people and implementing new forms of marketing for adaptation campaigns (Abed et al. 2015) have further argued that in fact, the way consumers communicate with each other has changed over the last decade, as retains the way they gather product information, exchange, get products and consume them, they have evolved into commercial platforms that more and more companies are utilizing. Social media is a source of business information through effective product development, marketing and distribution, customer communication, and also improving customer loyalty and relationships; social media is useful for building trust, reputation and brand image in B2C e-commerce(Makki and Chang 2015). According to(Rogers 1976) Proposed that time remains a crucial aspect in the adaptation process, the length of time it takes to decide to adopt innovation to use new ideas is also relevant when considering how an individual or other adoption unit changes its internal state (e.g. knowledge or decision to adopt) and its manifest behaviour (actual adoption or rejection)

- *H2a Social Media has a positive effect on B2C adaptation.*
- *H2b There is a significant and meaningful relationship between Infrastructures Innovation and interpersonal communication in the adaptation of B2C e-commerce.*
- *H2c Time has a constructive relationship with awarenessand infrastructurein an adaptation of B2C e-commerce.*
- *H2d Interpersonal Communicationhas a positive effect on B2C adaptation.*

*3.3 Infrastructure and Government policy*

Inadequate national and regional physical infrastructure, like roads, ports and broadband, as well as unreliable electricity supply, are serious barriers to e-commerce in Tanzania. To a large extent, these phenomena hamper the adaptation of B2C trade and place a nation at a disadvantage in global competition(Mlelwa, Chachage and Zaipuna 2015) At the same time (Awa et al. 2015)have argued that clear government policy on e-commerce is the first step towards adaptation,they argued more tax and tariff policies need to review, regulatory frameworks, subsidies and innovation infrastructure need to be well defined and promoted. Besides, inadequate infrastructure and interconnectivity do not exist in a vacuum; infrastructure like energy, software and hardware suppliers, skilled labour, broadband, backbone and fibre optics need to present for communities to integrate web-based services to help adaptation. (Laudon and Traver 2016) found that poor ICT infrastructure and shortage of technical and managerial skills are barriers to e-commerce adoption. (Rahayu and Day 2017) also revealed that poor communication infrastructure, lack of ICT knowledge, lack of IT resources, lack of financial resources and poor legal infrastructure are some factors leading to low ICT adoption in developing countries. (Choshin and Ghaffari 2017) Suggested that to expand the acceptance and adoption of e-commerce, it is essential to agree on the requirements of the technology, including telecommunications infrastructures, legal, security and messaging issues. The presence of a high-speed Internet, appropriate communications network, adequate organizational infrastructure, educational play a role in the success of B2C e-commerce. (Lawrence, Tar and journal 2010) suggested that before citizens and businesses in developing countries can consider participating in e-commerce, they must address the problems of lack of telephone lines, insufficient quality, slow and high cost of bandwidth and security. Most developing countries do not have ICT policies to guide the delivery of Internet services(Garg and Choeu 2015). No progress can be made without effective policies and resolute implementation of those policies. The lack of policies to guide the expansion of e-commerce in developing the country is a significant obstacle to the adoption of e-commerce.(Alyoubi 2015) Government initiatives are essential to the adoption of e-commerce and other ICTs. They may involve promoting the use of ICT, education and establishing an appropriate regulatory framework for e-commerce. Telephone access and competition among Internet service providers are key areas where government policies can influence access and adoption of e-commerce

- **H3a** *Infrastructure Innovation has a positive effect on B2C adaptation.*

-**H3b** *Time has a significant and meaningful relationship with Infrastructure Innovation in the adaptation of B2C e-commerce.*

-**H3c** *Government policies are affected overtime on the adaptation of B2C e-commerce.*

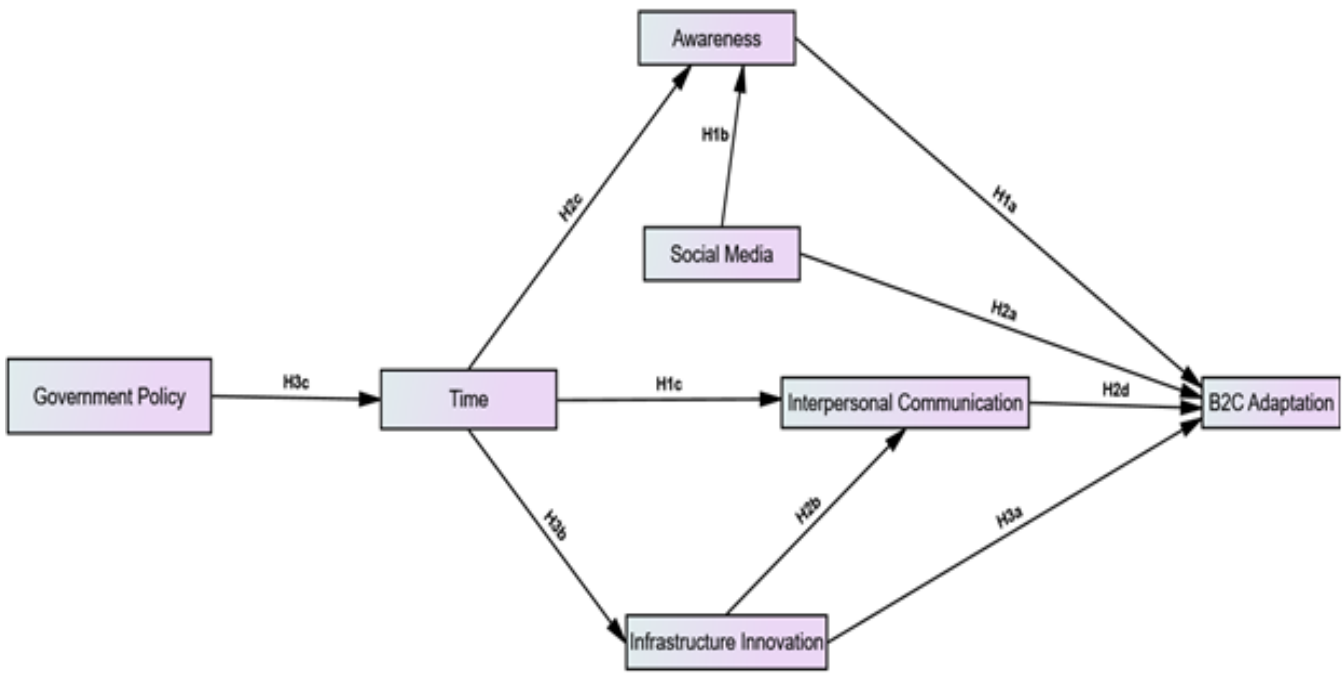

*Figure 1 Proposed research model*

## IV. RESEARCH METHODOLOGY DATA COLLECTION

　　　To test the hypotheses, a questionnaire was distributed to 279 respondents involving students, entrepreneurs, and workers of the public and private sector, According to (Kabir 2016)Sampling Is a process used in statistical analysis, in which a predetermined number of observations are taken from a larger population. The party chose to represent the population naturally became the sample sizes that a researcher uses to get empirical information or primary data. Moreover, we contacted participants via email, Facebook account and mobile phone. Some participants were selected from existing friends. The questionnaires were sent to those who were confirmed to participate through private emails and Facebook messages. Participants were humbly invited to respond to questions. Respondents who didn't, have experience in e-commerce, who heard about it and those who didn't respond to requests were excluded from the sample; participants were supposed to be frequent users of the Internet. Non-Internet users are excluded from the sample size. Second, according to (Datta 2011), Internet users do not always involve e-commerce users; participants who have answered imperfect questions related to the role of awareness, mass media and infrastructure innovation in adapting e-commerce has been excluded. The quantitative method has been used because it is objective, elaborate, and investigative (Sheard 2018)besides; the results obtained from the methodology are logical, statistical, and impartial. Data collection was collected in a structured manner on larger samples to represent the entire population. After eliminating 29 invalid responses, a total of 279 responses were used for analysis. Table 1, summarizes the profile of the 279 usable responses analyzed by participants. The gender of respondents was not relatively equal (65.6% male; 34.4% female). Most respondents were between the age of 26 and 35 (68.5%) with 91% under the age of 30.

**Table 1. Demographic characteristics of the respondents**

| No | Construct | Item | Frequency | Percentage % |
|---|---|---|---|---|
| 01 | Gender | Male | 183 | 65.6 |
|  |  | Female | 96 | 34.4 |
| 02 | Age | Under 25 years old | 36 | 12.9 |
|  |  | 26 to 35 years old | 191 | 68.5 |
|  |  | 36 to 45 years old | 43 | 15.4 |
|  |  | Over 45 years old | 9 | 3.2 |
| 03 | Education Level | PhD | 30 | 10.8 |
|  |  | Masters | 91 | 32.6 |
|  |  | Bachelor | 102 | 36.6 |
|  |  | Diploma | 42 | 15.1 |
|  |  | High School & Certificates | 8 | 2.9 |
|  |  | O-level & others | 6 | 2.2 |
| 04 | Occupation | Government employee | 87 | 31.2 |
|  |  | Private employee | 36 | 12.9 |
|  |  | Self-employed | 36 | 12.9 |
|  |  | Student | 76 | 27.2 |
|  |  | Unemployed | 12 | 4.3 |
|  |  | Entrepreneurs | 32 | 11.5 |
| 05 | Internet Frequency | Daily | 227 | 81.4 |
|  |  | Weekly | 40 | 14.3 |
|  |  | Monthly | 6 | 2.2 |
|  |  | Other | 2 | .7 |
|  |  | Not at all | 4 | 1.4 |

*4.1 Measurement development*

A positivist survey methodology has adopted. The online questionnaire was developed using Google forms. We adapt seven constructions from the previous research. All questionnaires were adopted based on existing research studies, as we have seen below. However, necessary changes had made to adjust the scope to the B2C e-commerce context. The measurement elements of each construction are listed in Schedule A. Measures of interpersonal communication, the decision to adopt B2C and the effect of time on adaptation were each assessed using the three-element scales adapted from(Chong et al. 2013),(Rahayu and Day 2017)adapted from(Jamal and Naser 2002), (Fang et al. 2014),(Alrousan and Jones 2016) and (Grandón, Nasco and Mykytyn Jr 2011)Government policy decision was evaluated using five scales adapted from (Tan et al. 2007),(Park and Kang 2014)(Awa et al. 2015)and (Faloye and Development 2014).Four elements to measure awareness through social media have also adapted from(Sun et al. 2012). we used four measurement elements from(Park and Kang 2014),(Tan et al. 2007),(Awa et al. 2015) and (Kurnia et al. 2015)to measure innovation in infrastructure. Social media was measured using a three-item scale adapted from(Kim, Ferrin and Rao 2009) and (Iankova et al. 2019) Finally, we included several demographic variables (i.e. gender, age, education level, occupation and Internet frequency as control variables in the model. To measure the magnitude of this study, we were reviewed and adjusted for previous studies to produce a questionnaire containing several 5-point Likert options, ranging from 1 - strongly disagree to 5 - strongly agree, Participants were asked to rate their level of dissatisfaction or satisfaction for each statement given under the seven concepts, namely interpersonal communication, the decision to adopt B2C, the effect of time on adaptation, a decision on government policies, awareness, innovation in infrastructure and social media. Appendix 1 shows the building questionnaires.

*4.2 Data analysis and results*

The data were analyzed using the Partial Least Squares (PLS) method, a structural equation modelling (SEM) technique based on path analysis and regression analysis, and recently, it has become increasingly popular for the analysis of multiple-construction relationships. Partial Least Squares (PLS) was adopted to perform data analysis because it is an established technique used in Structural Equation Modeling (SEM) to test causal relationships based on empirical data (Henseler 2017). According to (Hair et al. 1998) although there are still some precautions to be taken when using PLS, there are several advantages for choosing PLS as a statistical approach for modelling structural equations First, PLS path modelling analyzes a small sample well and avoids the problems associated with inappropriate sample size. Second, PLS allows more efficient analysis of complex structural equation models with many constructs and variables. Third, PLS requires fewer assumptions about the normal distribution of variables and error terms. Fourth, PLS simultaneously deals with both reflective and formative indicators. Fifth, PLS is better suited for theory development than theory testing and is particularly useful for forecasting. Sixth and finally, PLS overcomes the problems associated with many collars. Among the reasons above, we chose PLS as the method of analysis in this research.

We measured reliability using SPSS Statistics 23 to determine the correlation between the measurements items used to measure the latent variables in the model. The model was analyzed by using Cronbach's alpha (CA), composite reliability (CR), and the average variance extracted (AVE) values. Here, all factor loadings are greater than 0.71, which is satisfactory for the 0.5 cutoffs. Second, CA and CR values are higher than 0 .7(Bagozzi and Yi 1988) which have shown that the scale is consistent and reliable. Third, AVE values of all constructs are higher than 0.5 thresholds(Hair et al. 2011), which showed an adequate convergent validity of all constructs used in this research. These results are depicted in Table 2. Construct validity was tested using the subcategories of convergent validity and discriminant validity. Convergent validity was examined in terms of average variance extracted (AVE), and it was accepted when AVE was greater than 0.5. Discriminant validity was also examined in terms of AVE, and it was accepted when the square root of AVE was larger than all cases of correlations between each pair of constructs. Table 5 shows that all AVE were above 0.5, indicating that most of the variances were explained by their constructs.

**Table 2 Results of confirmatory factor analysis.**

| *Main Construct* | *Measurement Items* | *Factor Loading* | *Cronbach Alpha (CA)* | *Composite Reliability (CR)* | *Average Variance Extracted (AVE)* |
|---|---|---|---|---|---|
| Interpersonal Communication | InPC1 | .843 | .872 | 0.88781 | 0.725197 |
|  | InPC2 | .835 |  |  |  |
|  | InPC3 | .876 |  |  |  |
| The decision to Adapt B2C | DaB2C1 | .909 | .889 | 0.929438 | 0.814502 |
|  | DaB2C2 | .893 |  |  |  |
|  | DaB2C3 | .905 |  |  |  |
| Time affects Adaptation | TaA1 | .841 | .853 | 0.900506 | 0.751156 |
|  | TaA2 | .889 |  |  |  |
|  | TaA3 | .869 |  |  |  |
| Government Policy | Gp1 | .825 | .909 | 0.925203 | 0.713023 |
|  | Gp2 | .887 |  |  |  |
|  | Gp3 | .901 |  |  |  |
|  | Gp4 | .855 |  |  |  |
|  | Gp5 | .745 |  |  |  |
| Awareness | Awn1 | .802 | .769 | 0.849118 | 0.584791 |
|  | Awn2 | .761 |  |  |  |
|  | Awn3 | .761 |  |  |  |
|  | Awn4 | .732 |  |  |  |
| Infrastructure Innovation | InfrI1 | .802 | .916 | 0.849118 | 0.584791 |
|  | InfrI2 | .761 |  |  |  |
|  | InfrI3 | .761 |  |  |  |
|  | InfrI4 | .732 |  |  |  |
| Social Media | SoM1 | .855 | .803 | 0.879243 | 0.708335 |
|  | SoM2 | .811 |  |  |  |
|  | SoM3 | .857 |  |  |  |

*4.3 Direct Model Results*

We investigated whether the government policy to adopt B2C e-commerce is a significant impact over time and whether Awareness, Social media, Infrastructure Innovation and Interpersonal Communication are significant to the adaptation of B2C. Table 3 summarizes the path significance, coefficient and the results of the hypothesis tests. Besides describes the test results of the structural model path coefficients to test the proposed hypothesis. The first review of the relationship between the Gp and TaA, the results showed that H3c is supported (β =.0033; p <.05), the second relationship between the TaA and InfrI results showed that H3b is supported (β =.02; p <.05), while the direct path from TaA to Awn has not reached significance, with a path coefficient of 0.6 (p-value > 0.05), this implies that the H2C was not significant. It can be inferred that Length time has no significant influence on Awareness, fourth direct path of SoM to Awn achieved significance, with a path coefficient of 0.02 (p-value < 0.05), this implies that H1b was significant. Fifth both the direct path of InfrI to InPC and TaA to InPC achieved significance, with a path coefficient of *** (p-value < 0.001), this implies H2b and H1c were significant. Seventh the direct path of Awn to DaB2C failed to reach significance, with a path coefficient of 0.8 (p-value > 0.05) this implies H1a was not significant. Eighth both the direct path of InfrI to DaB2C and InPC to DaB2C have achieved significance with a path coefficient of *** (p-value < 0.001), and last the path of SoM to DaB2C did not reach significance, with a path coefficient of 0.5 (p-value > 0.05) this implies H1a was not significant. It can be inferred that Social Media does not have a significant influence on the adaptation of B2C.

Table 3 Results of the hypothesized effects.

| Hypothesis | Path | S.E. | C.R. | Path Coefficient β | Results |
|---|---|---|---|---|---|
| H3c | TaA <--- GP | 0.049 | 2.132 | 0.033 | Supported |
| H3b | InfrI <--- TaA | 0.083 | 2.331 | 0.02 | Supported |
| H2c | Awn <--- TaA | 0.07 | -0.51 | 0.61 | Not Supported |
| H1b | Awn <--- SoM | 0.07 | 2.294 | 0.022 | Supported |
| H2b | InPC <--- InfrI | 0.065 | 5.114 | *** | Supported |
| H1c | InPC <--- TaA | 0.084 | 4.479 | *** | Supported |
| H1a | DaB2C <--- Awn | 0.121 | 0.193 | 0.847 | Not Supported |
| H3a | DaB2C <--- InfrI | 0.051 | 8.355 | *** | Supported |
| H2d | DaB2C <--- InPC | 0.052 | 4.955 | *** | Supported |
| H2a | DaB2C <--- SoM | 0.108 | 0.662 | 0.508 | Not Supported |

Note: *Supported*at: *p < 0.05, **p < 0.01, ***p < 0.001, ns. = *Not Supported* (p > 0.05)

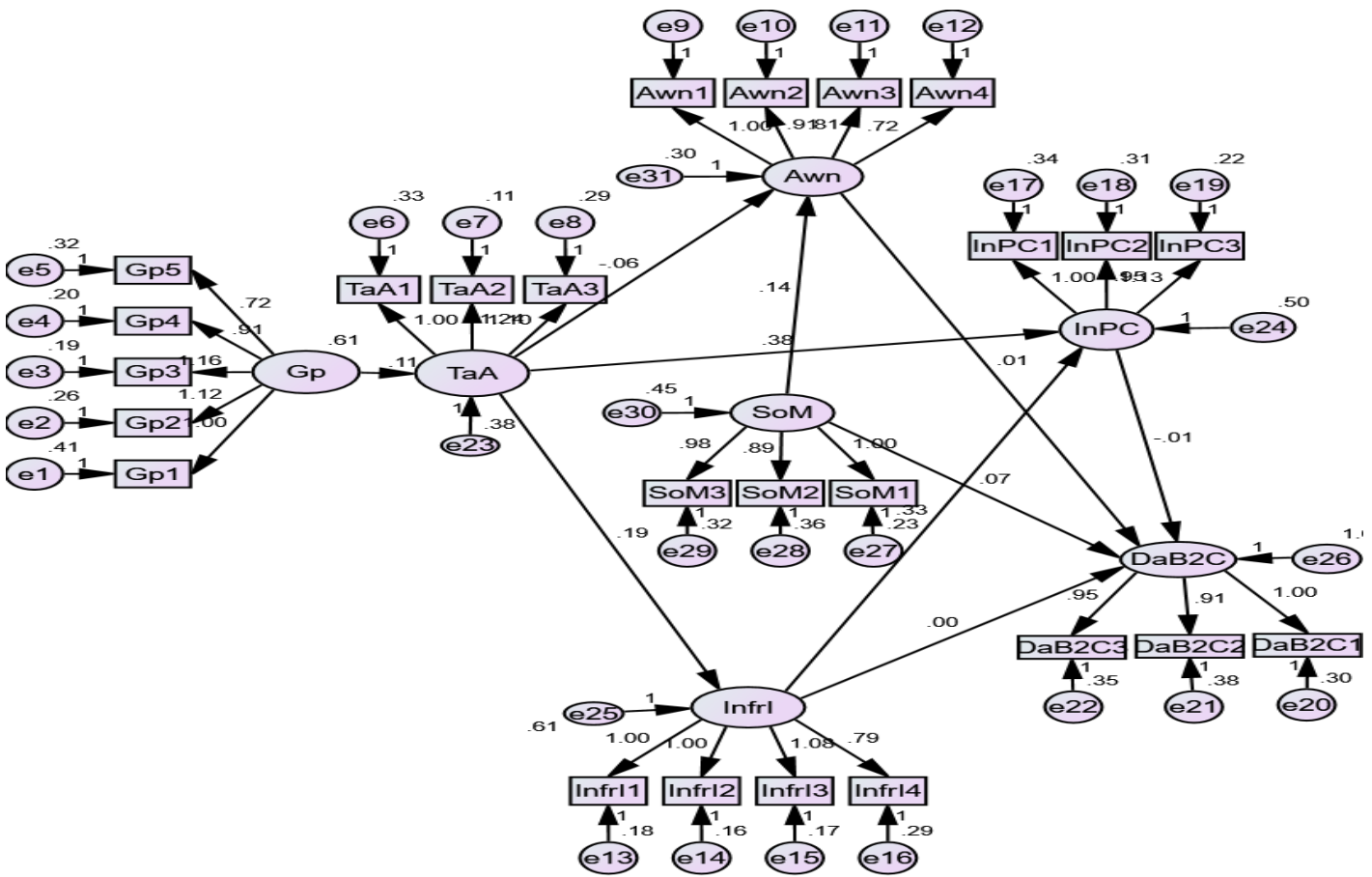

Figure 2 Results of structural equation model

Table 4, Mean Standard Deviation and Correlation.

| No | Variables | Mean | Std. Deviation | 1 | 2 | 3 | 4 | 5 | 6 | 7 |
|---|---|---|---|---|---|---|---|---|---|---|
| 1 | InPC | 12.0323 | 2.64895 | 1 | | | | | | |
| 2 | DaB2C | 11.5484 | 3.06001 | -.012 | 1 | | | | | |
| 3 | TaA | 12.3799 | 2.25667 | .290** | -.054 | 1 | | | | |
| 4 | GP | 19.7993 | 4.03080 | .305** | -.041 | .102 | 1 | | | |
| 5 | Awn | 16.7563 | 2.19009 | .034 | .014 | -.072 | .034 | 1 | | |
| 6 | InfrI | 16.4014 | 3.18779 | .338** | -.009 | .135* | .302** | -.051 | 1 | |
| 7 | SoM | 12.3190 | 2.14628 | -.140* | .041 | -.079 | .013 | .141* | -.040 | 1 |

**. Correlation is significant at the 0.01 level (2-tailed).

*. Correlation is significant at the 0.05 level (2-tailed).

## V. DISCUSSION AND CONCLUSION

We examined the role of awareness, social media and infrastructure innovation in adopting e-commerce. Using the innovation diffusion theory, the results indicated that when the government's use of social media at the right time to promote B2C e-commerce can encourage awareness and help adaptation, but also the government's policy to improving infrastructure innovation over time facilitates the adaptation of B2C e-commerce., this means that good telecommunications networks, security system and websites can increase people's confidence in B2C. Besides, Interpersonal communication has been affected by the length of time and infrastructure innovation in the adaptation process. This study provides a useful understanding of e-commerce adoption in developing countries, particularly Tanzania. The result shows that the adoption of B2C e-commerce in East Africa, particularly in Tanzania, is still at a low level. Most people are social media adapters. This condition certainly has implications for the government to redouble its efforts in raising awareness of the importance of electronic commerce for the development of the nation. Seminars, conferences or workshops should be held to provide opportunities for entrepreneurs, business stakeholders and people on e-commerce. Thus, awareness is one of the crucial issues affecting the development of e-commerce in East Africa. Social media has a significant impact on raising awareness and acts as an intermediary for the government to use and increase citizen awareness the lack of good policies discourages adoption. Also, income-related access fees affect Internet use. Monthly charges for Internet access are still very high in most developing countries. Because of inequalities in income distribution, the Internet is unaffordable for a large part of the population in rural areas. The dichotomy between urban and rural areas in term of technology use is a common feature of all developing countries. In urban areas, the use of ICT is quite common; while in rural areas of some developing countries, many small businesses do not even have computers yet, talk less about Internet access. There is a need for early computer education so that people can learn computer skills at school. It argued computer literacy populations had greater potential for appreciation and participation in B2C e-commerce. This study also provides empirical support that e-commerce offers many benefits for Tanzanians. Among the four benefits of B2C e-commerce reported by online zoom Tanzania include. Starting a website is cheaper at any time than a physical outlet. You don't have to provide your point of sale you don't have to pay rent and hire several employees to work there. The cost of marketing and promotion strategies is also low, flexible and fast - an individual or business can easily open an online store in a few days while a physical outlet needs space, commercial rental procedure as well as many constructions and decoration opportunities for its opening. Wide range of products and services — internet e-commerce allows customers to choose a product or service of their choice from any supplier anywhere in the world, saves time — the time taken to select purchase.

There are some limitations to this survey. First, the use of an online survey for data collection excluded people who did not have access to the Internet from participating in the study. Further research is needed to collect data using offline and online tools. Second, some people were not sufficiently aware of e-commerce because it is still at an early stage in Tanzania. Due to the delay in data collection, other researchers should raise awareness of the individuals who have targeted the sample size.Second, this survey examined the adaptation of B2C in the content of awareness, infrastructure and social media; further research may study the adaptation of B2C e-commerce based on the components of diffusion theory, which is knowledge, persuasion, the decision, implementation and confirmation. Third, the data from this survey were collected in Tanzania's largest city called Dares Salaam, therefore, further comparative case study research should target rural and remote areas with comparisons made to regions of major cities would be desirable. The future survey should explore in-depth the role of social media in raising awareness, people's culture and laws in the adoption of e-commerce. Finally, the online data collection survey was conducted in Tanzania. Although Tanzania is one of Africa's developing countries, the results of this study may differ from an East African country. Therefore, generalizations to other developing countries from this study should be cautious.

## ACKNOWLEDGEMENTS

Foremost, I would like to express my sincere gratitude to the Tanzania Ministry of Health, Community Development, Gender, Elderly and Children (Community Development) and the Government of the People's Republic of China for their continued support of my study and research; we also acknowledge the excellent comments and suggestions made by members of the review team.

**Appendix A.**

**Construct weight**　　　　　　**Measurement Item**

Interpersonal Communication　　　1, Do you agree that interpersonal communication in B2C e-commerce often cannot simplify adaptation?
2, Do you agree on interpersonal communication between people of strong ties can accelerate the adaptation of B2C e-commerce?
3, Do you agree that interpersonal communication can be used to improve the understanding of the peoples of B2C

The decision of B2C e-commerce　　4, When was the last time to buy items online
　　　　　　　　　　　　　　　　5, Have you ever used a website or applications to shop online?
　　　　　　　　　　　　　　　　6, I don't know much about how to shop on the Internet.

Adaptation of Time　　7, I don't think the time is taken into account when we divide the B2C adapter into different categories
 8, Do you think B2C adaptation will not be encouraged near future in Tanzania.
9, I believe that time is not the main factor in the occurrence of B2C adaptation.

Government Policy 10, I agree that the lack of support from the government agencies affect the adaptation of businesses to retail.
 11, I believe that if governments develop good strategic policy, raise awareness, build trust through social media and develop ICT infrastructure, they will promote B2C e-commerce.
 12, I believe that the government has taken sufficient measures (such as training, media advertising) to promote the
　　　　adaptation of B2C e-commerce.
13, I believe that governments should ensure that telecommunications services are modern and efficient in order to bring down prices for network use through effective. competition and market liberalization
14, I agree that the lack of a policy, regulatory and institutional framework hinders the adaptation of B2C e-commerce in Tanzania.

Awareness 20, I believe that little awareness about e-commerce is the reason for not adopting e-commerce.
21, Awareness is a direct influence of customer confidence in the adaptation of B2C e-commerce
22, I believe that less attraction to B2C e-commerce in Tanzania is caused by the lack of awareness and information about it.
23, I intend to use B2C e-commerce in the future if awareness is promoted, the ICT infrastructure is organizedand the e-commercepolicy is strengthened.

Infrastructure Innovation 24, I believe that the use of B2C e-commerce would improve business performance and the economy in the country if awareness promoted
25, I believe reliable infrastructure paves the way for faster adaptation of B2C e-commerce
26, I believe Tanzania has adequate ICTs such as software, computers, networks and corporate websites to support and encourage the adaptation of B2C e-commerce.
27, I believe that Tanzania's telecommunications infrastructure is unreliable and ineffective in supporting the adaptation of B2C e-commerce

Social Media 28, Do you believe that social media is a tool used to increase people's knowledge about adapting to e-commerce?
29, Do you agree on Interpersonal communication between people of strong ties can accelerate the adaptation of B2C e-commerce?
30, Information Communication Technologies (ICT) and related systemshave significant potential to help adapt B2C e-commerce and improve online business conditions.
29, Do you think the social media app can be used to encourage the adaptation of B2C e-commerce?
30, B2C-adapted ads on social media sites can significantly affect users' adoption intention